\documentclass[manuscript]{aastex}

\shorttitle{Hubble Space Telescope Observations of BALQSO Ton 34}
\shortauthors{Krongold et al.}

\begin{document}


\title{Hubble Space Telescope Observations of BALQSO Ton 34 Reveal a Connection between the Broad Line Region and the BAL Outflow\footnote{Based on observations made with the NASA/ESA Hubble Space Telescope for program HST-GO-12510, obtained at the Space Telescope Science Institute. STScI is operated by the Association of Universities for Research in Astronomy, Inc. under NASA contract NAS 5-26555}}


\author{Krongold, Y.\altaffilmark{1}, Binette, L.\altaffilmark{1},  Bohlin, R.\altaffilmark{2},  Bianchi, L.\altaffilmark{2}, Longinotti, A.L.\altaffilmark{3}, Mathur, S.\altaffilmark{4,5}, Nicastro, F.\altaffilmark{6,7,8},  Gupta, A.\altaffilmark{4}, Negrete, C.A.\altaffilmark{1}, Hernandez-Hibarra, F.\altaffilmark{9}}
\altaffiltext{1}{Instituto de Astronomia, Universidad Nacional Autonoma de Mexico, Apartado Postal 70264, 04510 CDMX, Mexico}
\altaffiltext{2}{Space Telescope Science Institute, 3700 San Martin Drive, Baltimore, MD 21218, USA}
\altaffiltext{3}{Catedrática CONACYT — Instituto Nacional de Astrofísica, Optica y Electronica, Luis E. Erro 1, Tonantzintla, Puebla, C.P. 72840, Mexico}
\altaffiltext{4}{Department of Astronomy, The Ohio State University 4055 McPherson Laboratory, 140 West 18th Avenue, Columbus, OH 43210-1173}
\altaffiltext{5}{Center for Cosmology and Astro-Particle Physics (CCAPP), 191 West Woodruff Avenue, Columbus, OH 43210, USA}
\altaffiltext{6}{Osservatorio Astronomico di Roma - INAF, Via di Frascati 33, I-00040 Monte Porzio Catone, RM, Italy}
\altaffiltext{7}{Harvard-Smithsonian Center for Astrophysics, 60 Garden St., MS-04, Cambridge, MA 02138, USA}
\altaffiltext{8}{University of Crete, Department of Physics, P.O Box 2208, GR-71003 Heraklion, Greece}
\altaffiltext{9}{Instituto Tecnológico Superior de Tequila (ITS-Tequila) Apartado Postal 46400 Tequila, Jalisco, México}






\begin{abstract}
Ton 34 recently transitioned from non-absorbing quasar into a BALQSO. Here, we report new HST-STIS observations of this quasar. Along with CIV absorption, we also detect absorption by NV+Ly$\alpha$ and  possibly O VI+Ly$\beta$. 
We follow the evolution of the CIV BAL, and find that, for the slower outflowing material, the absorption trough varies little (if at all) on a rest-frame timescale of $\sim2$ yr. However, we detect a strong deepening of the absorption in the gas moving at larger velocities ($-20,000$ -- $-23,000$ km s$^{-1}$). The data is consistent with a multistreaming flow crossing our line of sight to the source.
 The transverse velocity of the flow should be $\sim$ few thousand km s$^{-1}$, similar to the rotation velocity of the BLR gas ($\approx 2,600$ km s$^{-1}$). By simply assuming Keplerian motion, these two components must have similar locations,  pointing to a common outflow forming the BLR and the BAL.  We speculate that BALs, mini-BALs, and NALs, are part of a common, ubiquitous,  accretion-disk outflow in AGN, but become observable depending on the viewing angle towards the flow. The absorption troughs suggests a wind covering only $\sim$20 \% of the emitting source, implying a maximum size of 10$^{-3}$ pc for the clouds forming the BAL/BLR medium. This is consistent with constraints of the BLR clouds from X-ray occultations. Finally, we suggest that the low excitation broad emission lines detected in the spectra of this source lie beyond the wind, and this gas is probably excited by the shock of the BAL wind with the surrounding medium.    

\end{abstract}





\section{Introduction}

Outflows are ubiquitous in Active Galactic Nuclei (AGN). They manifest themselves through blueshifted absorption lines imprinted in the UV and X-ray continuum of AGN spectra (e.g. Kraemer \& Crenshaw 2003, Krongold et al 2003, 2009, Andrade-Velazquez et al. 2010, Gupta et al 2013a). They have been classified as having Narrow Absorption Lines (NALs), Broad Absorption Lines (BALs), or intermediate cases (mini-BAL, e.g. Hamman and Sabra 2004). 

These outflows are thought to be strong driving agents of galaxy evolution, producing the AGN feedback  required to stop the growth of their host galaxies, heating their interstellar medium and quenching star formation (di Matteo et al. 2005, Silk 2011, Kormendy \& Ho 2013). They may well be responsible, among other things,  for the well known relation between the central black hole mass of a galaxy and the velocity dispersion of its bulge, as well as for the steep decline of the number density of galaxies at high masses (Scannapieco \& Oh 2004; Hopkins  et al. 2005; King 2010; Ostrikerer et al. 2010). 

However, the real effect of these winds on their environments (if any) is not yet well established. There are some examples of outflows
capable of feedback. Some of these include BAL winds (Moe et al. 2009, Dunn et al. 2010,  Borguet et al. 2013, Capellupo et al. 2014) or NAL winds (Arav et al. 2013), where the kinetic outflow rates represent up to a few percent of the bolometric luminosities of the system. The recently discovered X-ray ultra-fast outflow winds (X-ray detected winds with outflow velocities larger than 10,000 km s$^{-1}$) also have high enough mass and kinetic outflow rates in order to produce feedback (e.g. Gupta et al. 2015, 2013b, Tombesi et al. 2015, Nardini et al. 2015, Longinotti et al. 2015). However, the role of AGN winds in galaxy evolution has not yet been proven. The outflow velocities are comparable to the escape velocities at the suggested outflow locations, making a large scale effect possible. But there are still large uncertainties in the mass and energy outflow rates of these winds, which are strongly dependent on the density of the absorbing gas and its distance to the central source, quantities that are difficult to estimate (Krongold et al. 2007, 2010). Even less is known on the physical mechanisms of their interplay with the interstellar medium of their host galaxies. Thus, in order to establish the cosmological effect of AGN winds, it is mandatory to understand their nature.

It has  been  suggested  that  the  different  outflow manifestations (NALs, BALs; mini-BALs) observed in the UV and X-rays are the result of a single large scale wind viewed at different angles (Elvis 2000; Gibson et al. 2009; Hamman et al. 2012; Giustini 2015, Matthews et al. 2016). Such wind would be ubiquitous in AGN but only visible when the gas crosses our line of sight (Goodrich \& Miller 1995; Hines \& Wills 1995; Murray et al. 1995; Elvis 2000; Krongold et al 2005). It has also been suggested that the same gas may be  responsible  for  absorption  and  emission  (e.g.,  Mathur et al. 1995;  Elvis 2000; Fields et al. 2005; Krongold et al. 2007; Andrade-Velazquez et al. 2010; Longinotti et al. 2013; Miniutti et al. 2014; Sanfrutos et al. 2016), or that BALs may be associated to the obscuring dusty torus (Leighly et al. 2015). However, other scenarios suggest that the incidence of BALs may be related to evolutionary phases (Hazard et al. 1984; Becker et al. 2000) and that mini-BALs and BALs may represent the same outflow but at different evolutionary stages (Hamann \& Sabra 2004). This lack of consensus evidences the need for more robust studies of AGN outflows.

Among these studies, perhaps the most enlightening ones are those focused on variability of the absorption lines. This is because these variations have strong diagnostic potential on the physical properties, location, and geometry of the associated winds, and as such, give insights on their potential capability of feedback (Krongold et al. 2007). Evidence of variations on BAL throughs has been well established over systematic studies in the last ten years (Barlow1994, Lundgren et al. 2007; Gibson et al. 2008, 2010; Filiz Ak et al. 2012, 2013; Capellupo et al. 2011, 2012, 2014; Wildry, Goad \& Allen 2013; Welling et al. 2014). However, examples of dramatic variability, where entire absorption troughs emerge or disappear are more scarce and just beginning to accumulate (Junkkarinen et al. 2001; Lundgren et al. 2007; Hall et al. 2011; Vivek et al. 2012a,b , 2015; Filiz Ak et al. 2012; Ma 2002; Hamann et al. 2008; Leighly et al. 2009, 2015; Krongold et al. 2010; Rodriguez Hidalgo et al. 2011, 2013, Chen et al. 2013, 2015). From the theoretical point of view, Hydrodynamical simulations predict strong time variability of the flows (Proga et al. 2012) 

In the literature, there are two different scenarios to explain the observed variations. The first consists on bulk motions of multistreaming gas across the sightline to the emitting source. The second involves changes in the ionization state of the gas, following variations of the photoionizing  impinging flux produced by the AGN. Variations in existing troughs have been explained with both scenarios. For instance, cases where variations in the observed flux and in the absorption lines are correlated, are consistent with changes in the ionization state of gas (e.g. Wang et al. 2015, Misawa et al. 2014). Cases where the trough varies simultaneously at a wide range of outflow velocities can also be explained by this scenario (Hamman et al. 2011, Capellupo et al. 2012), even if there is no information on the flux variability of the central source. This is because flux changes would modify the ionization state of all the gas along the line of sight (independently of the velocity of the gas), while bulk motions of material would produce variations only at the velocities where the motion of gas is taking place. In other cases, the variations are more easily understood in terms of material crossing into our line of sight, either because only portions of the trough vary (e.g. Capellupo et al 2012, 2014, Filiz Ak et al. 2013, Vivek et al. 2016) or because the troughs are strongly saturated and thus are insensitive to flux variations (e.g. Hamman et al 2008, Capellupo et al. 2009, McGraw et al. 2015). Most events of dramatic variability, where transient troughs appear or disappear, have been interpreted as motions of multistreaming gas across the sightline to the emitting source (Ma 2002, Hamman et al. 2008, Leighly et al. 2009, 2015, Krongold et al. 2010, Hall et al. 2011, Rodriguez Hidalgo et al. 2011, Chen et al. 2013, Longinotti et al. 2013, Kaastra et al. 2014; although see Chen et al. 2015).

Ton 34, at z$_q$ = 1.928 is one of those few objects where a dramatic transition has been observed, it changed from non-absorption quasar into BALQSO. Krongold et al. (2010, hereafter K10) detected an ultraviolet C~IV broad absorption trough, in a spectrum obtained in 2006 by the Sloan Digital Sky Survey (SDSS). However, two different spectra acquired in 1981 at Las Campanas and Palomar observatories (Sargent et al. 1988) did not show this absorption feature, evidencing the emergence of the outflow in less than $\sim$ 8 yr (rest-frame). This led K10 to suggest that the absorption trough, which spans a velocity range from $\sim$ -5,000 to -25,000 km $^{-1}$, is likely the result of an existing wind moving towards our line of sight (instead for instance, being due to the rise of a new wind or an existing wind that becomes visible in the UV because of changes in ionization). 

Ton 34 has shown other extreme properties in addition to the emergence of the BAL C~IV outflow. It presents a dramatic decline in the continuum shortward of $\approx~1100$ \AA, the so called UV-break in quasar spectra. Binette \& Krongold (2008a) found that moderate columns of carbon crystalline dust absorption (Binette et al. 2005, Haro-Corzo et al. 2007) could produce this particular shape of the SED. These authors further analyzed the
emission line spectrum of this quasar and suggested an unusual strength (relative to Ly$\alpha$) of low to intermediate excitation emission lines (such as OII + OIII $\lambda\lambda$835, NIII + OIII $\lambda\lambda$686, 703, and NIII + NIV $\lambda\lambda$765), not compatible with photoionization of the gas, but rather with shock excitation processes (Binette \& Krongold 2008b) .

In this paper we present new Hubble Space Telescope (HST) observations of quasar Ton 34 and confirm the C IV BAL trough reported by K10, as well as detection of Ly$\alpha$ + N V and Ly$\beta$ + O VI BAL absorption. We further report the rapid variability of the C~IV absorption feature in less than two years. The paper is organized as follows. In \S \ref{data} the data processing is described. In \S\ref{lya1} and \S\ref{c4} we present the spectral analysis and compare the observed spectra with previous data. Finally, in \S \ref{conc} we discuss the implications of our findings. Throughout the paper we assume a cosmology consisting of H$_o=70$ km s$^{-1}$ Mpc$^{-1}$, $\Omega_M = 0.25$, and
$\Omega_\lambda = 0.75$.

\section{HST-STIS Observations of Quasar Ton 34 \label{data}}


We observed Ton 34 with the Cosmic Origins Spectrograph (COS) and Space
Telescope Imaging Spectrograph (STIS) in 2012 through Proposal ID 12510,  PI L. Binette. COS data will be presented in a forthcoming paper studying the broad-band X-ray to UV SED of Ton 34. Here we focus on data obtained by STIS with gratings G230L and G430L using the 52$\times$0.1 arcsec aperture. These data fully cover the wavelength range between 1685 and 5700 \AA, and are ideal to study the Lyman and CIV region around 1000-1600 \AA\ in the rest-frame of Ton 34. G430L observations were carried out on March 28, 2012 , and had a duration of $\approx2000$ s. G230L data were acquired on May 20, 2012, for a total  of $\approx1900$ s.

The STIS data were reduced with a set of IDL routines written by D. Lindler for the STIS team and accessible to some of the authors for custom data reduction. Hot pixels in the CCD observations are flagged and repaired by linear interpolation along the dispersion direction. Cosmic ray hits are removed from the dithered CCD data using a four sigma rejection criterion. The wavelength scales agree with the standard STScI pipeline *\_x1d.fits files to 0.1 of a resolution element of two pixels. The flux accuracy is limited to  about 10\% by the poor repeatability of observations in the narrow 0.1" slit (Bohlin \& Hartig 1998). Actual variation among the four dithered observations is 2\% for G230L and 12\% for G430L.

\section{Absorption Troughs in the Spectra of Ton 34 \label{lya1}}

\begin{figure}
\plotone{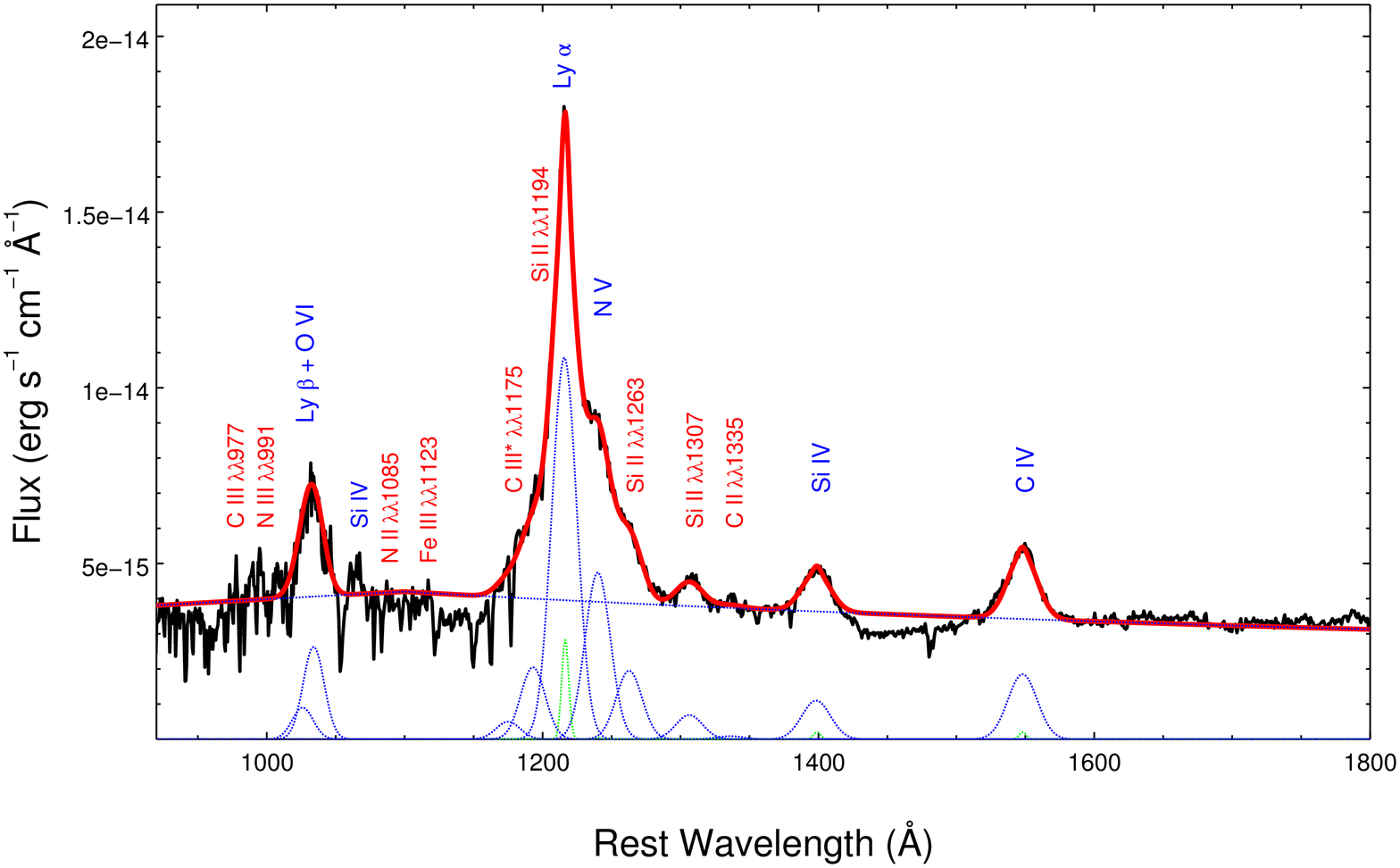}
\caption{HST-STIS spectra and model of quasar TON 34. Typical errors in the spectrum are $\approx2\%$ for the CIV NV+Ly$\alpha$ region and $\approx5\%$ for the OVI+Ly$\beta$ region. The continuum is modelled with a broken powerlaw at $\sim1100$\AA\ (blue dotted line). Emission features are fit with a narrow and broad Gaussian components (green and blue dotted Gaussians). Possible low excitation lines are also marked in red. Clear absorption troughs can be observed near 1450, 1150, and possibly at 960\AA, consistent with absorption by C IV, NV+Ly$\alpha$, and OVI+Ly$\beta$. \label{lya}}
\end{figure}

In Figure \ref{lya} we present the HST-STIS data in the rest-frame of Ton 34. The plot presents the data in the 920 --  1800\AA, range. The presence of the strong Ly$\alpha$ + N V, Ly$\beta$ + O VI, C IV, Si IV, and Si II $\lambda\lambda$1307 emission lines is evident in the spectrum (there might also be weak emission by C II $\lambda\lambda$1335). We fit the data with a broken powerlaw at 1100\AA. Around this region there is a change of slope in the continuum, consistent with findings by previous studies of this source (Binette \& Krongold 2008a). Each emission line was modeled using two Gaussian profiles, one to account for the possible presence of a narrow emission component, and another to account for the broad emission component. We constrained the full width half maximum (FWHM) of the narrow and broad components to be the same in all lines. With this approach, we find that most lines could be fit very well, with the positions of the lines consistent with being at rest in Ton 34 . However, 
the Ly$\alpha$-N V complex requires special consideration as it has a strong redward asymmetry, along with a weaker blue wing. Similar profiles have been observed in composite spectra of quasars (e.g. Suzuki 2006), and can be explained by the presence of Si II $\lambda\lambda$1194, and  $\lambda\lambda$1263 emission lines, which can be photo-pumped, and thus stronger than low-excitation lines (e.g.  Baldwin et al. 1996, Leighly et al. 2007).
We find that the full emission Ly$\alpha$-N V complex can be explained including emission by these two lines plus C III* $\lambda\lambda$1175 (e.g. Laor et al. 1997, Leighly et al. 2007), with positions consistent with the rest-frame of Ton 34 and FWHMs tied to the value obtained in modeling the other Broad Emission lines. Table 1 presents the best fit parameters of the emission lines used to model the spectra.

The best fit model is presented in Figure \ref{lya}, along with the different contributions from the narrow and broad components to the emission lines. While the model fits well most of the features in the spectral range between  920 --  1800\AA, there is clear underlying broad absorption near 1450\AA, that confirms the presence of the C IV BAL trough detected in this source by K10. We note that the data do not show the presence of absorption by Si IV.

In addition, there is also strong broad absorption near 1150\AA, that we identify with a BAL trough produced by a blend of NV and Ly$\alpha$, as commonly observed in BAL flows (e.g. Baskin, Laor \& Hamman 2013). We do not attempt to separate the contribution of each ion, but regard it as a combination of both, NV+Ly$\alpha$.   The data further show possible absorption near 960 \AA, consistent with a blend of OVI+Ly$\beta$. The presence of the absorption troughs is clearly independent of our modeling of the emission lines, as they are all observed below the continuum level  (see blue dotted line in Fig. \ref{lya}).

\begin{figure}
\plotone{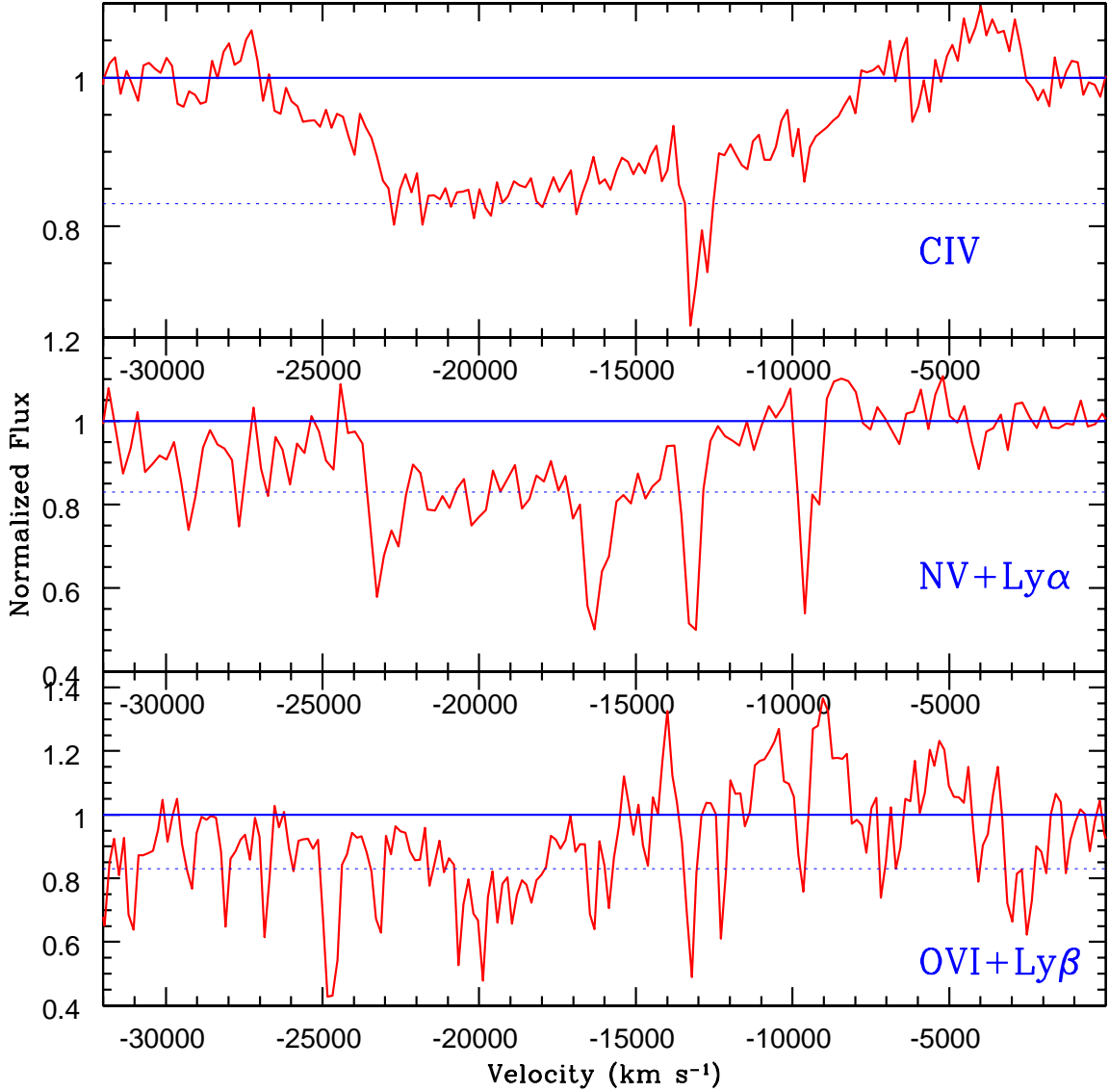}
\caption{Normalized HST-STIS spectra near the CIV, NV+Ly$\alpha$, and OVI+Ly$\beta$, presented in velocity space. In the middle and lower panel, the velocity is calculated with respect to the rest frame of the respective H transition. There is clear absorption by  CIV and NV+Ly$\alpha$, and possible absorption by OVI+Ly$\beta$ between -5,000 and  $-25,000$ km s$^{-1}$.  The blue dotted lines mark the maximum depth of all the absorption troughs, all consistent with a covering factor of the wind $\sim0.2$. The absorption by OVI+Ly$\beta$, if real,  may be filled by emission from C III $\lambda\lambda$ 977 and NIII $\lambda\lambda$ 991 low excitation gas (see Fig. 1), whose position in velocity in the lower panel would be in the range between -7,000 and -16,000 km s$^{-1}$. \label{vel}}
\end{figure}

Figure \ref{vel} presents the normalized spectra (the ratio between the data and the best fit emission model) in velocity space for each absorption trough. The upper panel shows that the CIV absorption extends from velocities $\approx$ -8,000 km s$^{-1}$ up to velocities $\approx$ -25,000 km s$^{-1}$. The middle panel shows the NV+Ly$\alpha$ trough in the Ly$\alpha$ rest frame. It can be observed that the absorption extends roughly through the same velocity range of the CIV (the NV+Ly$\alpha$ absorption extends from $\approx$ -12,000 -- -25,000 km s$^{-1}$, but there might be absorption at lower velocities filled by the emission lines, see below). In particular, we note the similarity in the absorption troughs at the high outflow velocity end. In this region, the Ly$\alpha$ transition should dominate the absorption, given the $\approx -5,900$ km s$^{-1}$ velocity difference with respect to the NV transition. Both, the C IV and Ly$\alpha$ troughs decline near -25,000 km s$^{-1}$. This supports the idea that there is a significant contribution of the Ly$\alpha$ opacity in this feature.

The lower panel of Figure \ref{vel} presents the  OVI+Ly$\beta$ region in velocity space, relative to the Ly$\beta$ rest frame. The data is suggestive of absorption, however,  the profile is different than the one observed in C IV and NV+Ly$\alpha$. In particular, the profile is much less broad than for the other two troughs, with lack of absorption between -7,000 and -16,000 km s$^{-1}$. These velocities correspond exactly to the spectral region where Binette \& Krongold (2008b) detected in 1981 Palomar data (see \S \ref{c4}) the presence of strong low excitation emission lines by CIII $\lambda\lambda$ 977 and NIII $\lambda\lambda$ 991. Indeed, there is evidence of these  lines in the spectrum, and several other low excitation emission lines are also present in the STIS (see Figure 1) and COS spectra (a detailed analysis of these lines will be presented in a forthcoming paper). Therefore, the lack of absorption in these ranges could be easily understood  by ``filling" of the absorption troughs by the CIII and NIII emission lines.

Interestingly, the residual normalized flux is roughly the same below the CIV and the NV+Ly$\alpha$ absorption troughs (see blue dashed line in Figure \ref{vel}), corresponding to a value  $\sim$0.8. The OVI+Ly$\beta$ trough, reaches also this residual normalized flux, in the regions where no contamination is expected by the possible presence of the CIII and NIII emission lines (i.e. outside of the region between -7,000 and -16,000 km s$^{-1}$). The similarities in the profiles of the troughs suggest that the lines are saturated, and their maximum depths do not depend 
on the total column density of the gas, but rather on the fraction of the gas covering the emitting source. This has been observed in other outflows (e.g., Arav et al. 2001; Hamann \& Sabra 2004; Gabel et al. 2005). If the troughs were not saturated in the spectrum of Ton 34, one would expect that each of them had a residual flux determined by its own opacity. Given that there is absorption by both  NV and Ly$\alpha$ in the 1150\AA\ trough, one would expect a much larger opacity in this region than in the CIV region. This has been observed in the average spectrum of BALs (see Figure 2 in Baskin, Laor \& Hamman 2013). Therefore, the simplest interpretation of this result is that indeed the troughs are saturated, and the maximum depth is indicative of the covering factor of the flow. Then the outflow covering factor should be small, $\sim$0.2.

\section{The Evolution of the CIV Absorption Trough \label{c4}}

We studied the evolution of the CIV trough by comparing the HST-STIS data with those obtained by the Sloan Digital Sky Survey (SDSS) on January 30, 2006 (see details in K10), and by the Palomar 5.08m Hale Telescope, between November 19-21, 1981 (Sargent et al. 1988).  To assess the possible presence of opacity variations in the absorption trough, we fit the 2012 spectrum following the same approach used with the 2006 and 1981 data (K10): We fit a simple powerlaw to the line free regions in the spectrum, between  3000-3100 \AA~(assumed to be a good region to measure the continuum  given the lack of strong Fe II emission) and 1320-1380\AA. As noted by K10, this approach is the most conservative to establish the presence of the absorption, and we follow it here to have a reliable comparison among datasets.  As in K10, the Si IV and C IV emission lines in the spectrum were fit using two different Gaussians for each emission line, to account for the narrow and broad emission components. This self-consistent strategy used to fit all the data make the comparison robust.

\begin{figure}
\plotone{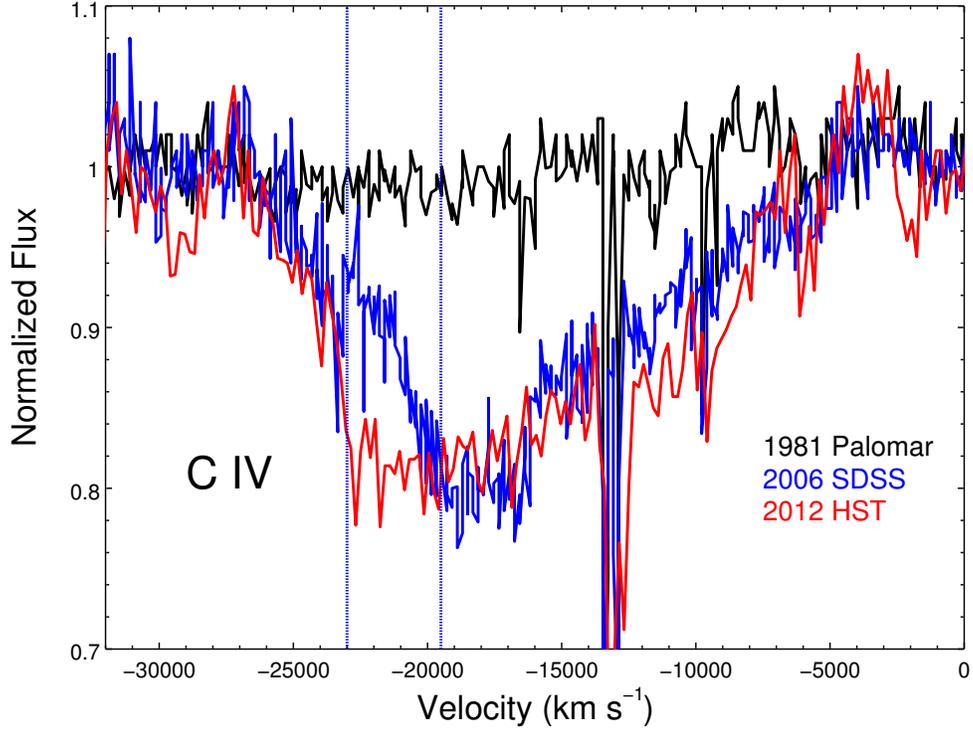}
\caption{Normalized spectra of Ton 34  from the Palomar Observatory, from the SDSS and from HST-STIS, near the CIV region. Typical errors are $\approx2\%$ for the HST, $\approx5\%$ for the SDSS, and $\approx1\%$ for the Palomar (Sargent et al. 1988) data. The data have been modelled using a self-consistent approach for the three datasets, and is presented in velocity space. The absorption trough was not present in 1981, yet it is clear during the 2006 and 2012 observations. While the absorption trough is similar between these later two spectra (perhaps with minor variations), it deepens at the large outflow velocity end of the trough (velocities $\sim$20,000 -- 23,000 km s$^{-1}$). This is indicated by the blue vertical dotted lines. \label{civ}}
\end{figure}

Figure \ref{civ} presents the normalized data in velocity space.  The profile of the C IV absorption feature is similar in the SDSS 2006 and STIS 2012 data for velocities lower than $\sim$-20,000 km s$^{-1}$, and larger than $\sim$-23,000~km s$^{-1}$.  A slight strengthening of the trough can be observed in the low velocity end of the trough, which is significant given the errors in the data (typical errors in the STIS data are of the order of 2\%, and in the SDSS data of 5\%, and of 1\% in the Palomar data (Sargent et al. 1988) near the CIV region). However, this strengthening could also be the result of observational uncertainties (for instance, continuum placement). We note however, that between -20,000 and -23,000~km~s$^{-1}$, there is a strong deepening of the absorption trough that took place between the 2.1 years separating the observations (rest-frame). 


\section{Discussion and Summary \label{conc}}

Ton 34 has transitioned from a non-BAL quasar into a BAL object in less than 8 yr (rest frame). The CIV trough seems to be evolving on a rather fast timescale, less than 2.1 yr  (in the rest frame of the source, herafter t$_{CIV}<$2.1 yr), according to a straightforward comparison between the SDSS data and the STIS data. In particular, the high velocity end of the absorption (v$_{max}>-$20,000 km s$^{-1}$) has shown a strong deepening, reaching the same depth as the absorption trough observed in 2006 at larger velocities. In Ton 34, the throughs appear to be rather shallow, with the CIV trough presenting a low balnicity index\footnote{The balnicity index (which is to be applied to C IV) measures the equivalent
width of strong absorption features (expressed in km s$^{-1}$), but requires that
each absorption feature contributing to the index spans at least 2000 km s−1
(to exclude intervening systems) and only includes in the calculation
absorption regions dipping 10\% or more below the normalized continuum. In
addition, the first 3000 km s−1 blueward of the emission peak are excluded to
distinguish “associated absorption” from broad absorption. For further details
see Appendix A in Weymann et al. (1991).} (Weymann  et  al., 1991), $\gtrsim 800$  km s$^{-1}$. For comparison, typical BALQSOs seem to have much deeper troughs, with CIV balnicity much larger than $\sim 2,000$  km s$^{-1}$ (e.g. Weymann et al. 1991, Gibson et al. 2009). Ton 34 is also on the high end of FWHM in the absorption trough, we measure a FWHM$\sim 10,000$ km s$^{-1}$. Baskin et al. (2015) found that only 4\% of BAL quasars have such large widths in the troughs. Transient absorption features in other sources have also shallow troughs, seem to evolve on timescales larger than a few years, and present more dramatic variability at larger outflow velocities (e.g. Filiz Ak et al. 2012, 2013, Vivek et al. 2016). Thus, Ton 34 fits well within this pattern.

\begin{figure}
\plotone{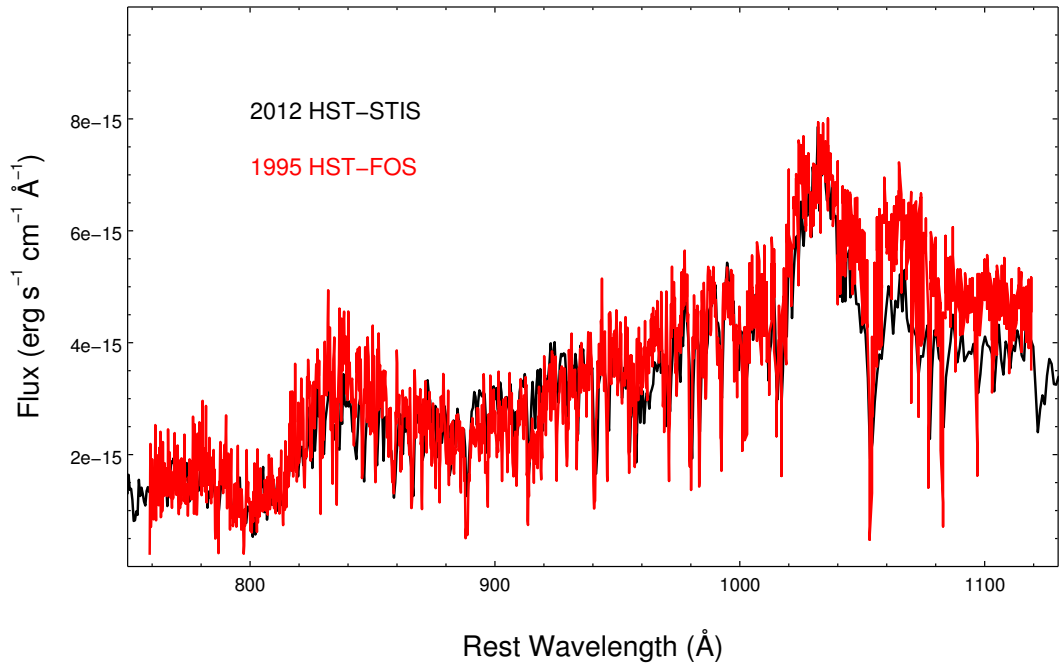}
\caption{Comparison of the spectra of Ton 34  taken by HST-STIS in 2012, and the one obtained by HST-FOS in 1995 (see details of this dataset in Binette \& Krongold 2008a). In the $\sim 6$ yr (rest-frame) ellapsed between the two observations the flux level of the source is similar.\label{broad}}
\end{figure}

The CIV trough variations in the multi-epoch spectra of Ton 34 might be related to ionization changes in the flow or to the transverse motion of a multistreaming flow into our line of sight (hereafter l.o.s.). Using photoionization equilibrium time-scales (Nicastro et al. 1999, Krongold et al. 2007, D'Elia et al 2009, Krongold \& Prochaska 2013), any gas recombining in less than t$_{CIV}<$2.1 yr sets a very loose lower limit on the number density of the outflow at n$>1000$ cm$^{-3}$. Since the flow is likely located close to the Broad Line Region (BLR, see K10 and argument below) and since the gas density in this region is expected to be orders of magnitude larger, then, the gas would have had time to react to flux changes and reach ionization equilibrium with the impinging ionizing continuum. However, there is no evidence of changes in the ionizing flux level in the extreme UV spectral region of Ton 34 over the last $\sim$6 yr (rest frame). Figure \ref{broad} shows that the 2012 HST-STIS data  have an ionizing continuum flux similar to the one measured via the HST-FOS observations carried out in 1995  (see Binette \& Krongold 2008a for details on the FOS data). In addition, apart from the deepening at $\sim -22,000$ km s$^{-1}$, the C IV trough shows only small variations (if any) between 2006 and 2012. This rules out the possibility that changes in the ionization level of the gas caused the observed absorption trough variations, and suggest bulk motions of gas crossing our line of sight.

The transverse velocity of the gas can be constrained using the simple relation   v$_{tr}\sim$R$_{em}$/t$_{cross}$ where R$_{em}$ is the size of the  continuum source emitting region,  and t$_{cross}$ is the crossing time. K10 estimated R$_{em}$(1500 \AA)$\sim few\times 10^{16}$cm, 
using arguments of flux variability studies in other quasars (Kaspi et al. 2007). The crossing time must be shorter than the time elapsed between the observations, thus t$_{cross}=$~t$_{CIV}~\lesssim$ 2.1 yr. Therefore, the transeverse velocity of the C IV absorbing material v$_{tr}\sim$~few thousand  km s$^{-1}$. This velocity is similar to the rotational velocity of the gas emitting in the C~IV Broad Line Region, v$_{BLR}\sim 2600$ km s$^{-1}$ (assuming the rotational speed of the gas as half the FWHM of the emission line, Table 1). By simply assuming Keplerian motion in the black hole potential, this means that the the location of both components should be the similar (as R$\propto$v$^{2}$). Thus, the observed variability in the C~IV absorption trough of Ton 34 strongly suggests a connection between the emitting and the absorbing gas. Using equation (3) from Kaspi et al. (2007), and the 1350~\AA\ luminosity of Ton 34, K10 estimated a  C IV Broad Line Region radius of R$_{BLR}\sim 2\times10^{18}$cm. Thus, the location of the absorption/emission system is $\sim$1pc.

We further stress that the rapid variability of the C IV trough rules out a large scale location (kpc from the central source) for the wind, as suggested for other BALs (Arav et al. 2013). Already at 10 pc away from the central engine v$_{cross}<1000$ km s$^{-1}$, and the crossing time of the wind into our l.o.s. would have to be $\sim$10 yr, much larger than t$_{CIV}$.

Of particular interest is the possible finding that the wind does not fully cover the emitting region, as we observe a systematic residual flux in the absorption troughs that suggests a covering factor f$\sim 0.2$. If the wind is part of the BLR, this would imply that it must be formed by a large number of clouds. We can put an upper limit to the transverse size of the clouds, simply as R$_{cl}~<$~R$_{em}$$\sqrt{\rm{f}}$, where R$_{em}$(1500 \AA)$\sim few\times 10^{16}$cm is the size of the emitting region estimated above, and f the covering factor. This implies R$_{cl}~<10^{15}$ cm, which is fully consistent with the constraints on the size of the BLR clouds inferred from X-ray occultations, i.e. R$_{cl}~<~few \times 10^{12}$ cm (Maiolino et al. 2010; Risaliti et al. 2011, see also Williams et al. 2017). 

We conclude that Ton 34 represents a bona-fide example of a quasar wind that is crossing our line of sight  to the continuum source (as suggested for other objects, e.g Ma 2002, Hamman et al. 2008, Leighly et al. 2009, Rodriguez Hidalgo et al. 2011). In the case of Ton 34, the parcels of the wind with larger radial speed seem to be crossing our line of sight with a lag with respect to the slower material.  

In this sense, the emergence of BAL flows after the appearance of mini-BALs may not be related to the evolutionary state of quasars (e.g. Hamann \& Sabra 2004). We speculate that the same wind is capable of producing both phenomenologies, with a mini-BAL transitioning to a BAL as the regions of the wind with larger velocity begin to cross our line of sight towards the emitting source. Indeed, Rodriguez-Hidalgo et al. 2011) found that quasar J115122.14+020426.3 transitioned from mini-BAL into BAL. Given that there is a strong perpendicular component in the velocity of the wind with respect to our l.o.s., a possible connection to the NAL phenomenology is also possible. In this case, the NAL would be produced in cases when most of the flow is seen in the transverse direction  (Elvis 2000, Krongold et al. 2007, Krongold et al. 2010; Hamman et al. 2013). Ton 34, thus, points towards an ubiquity of quasar winds around their central engines, but only becoming visible when they are located in our direct view to the continuum source.

Finally, we note that if the deepening near 960 \AA\ is indeed produced by absorption due to OVI+Ly$\beta$, and if our interpretation of the absorption profile is correct, the low excitation emitting gas may indeed be filling this trough (in particular by the C III $\lambda\lambda$ 977 and NIII $\lambda\lambda$ 991 emission lines). This would mean that the emitting gas with lower excitation should be located outside of the wind (as these lines do not seem to be absorbed). The low excitation material could be photoionized gas, whose low ionization level would be the result of the shielding of the central engine radiation by the fast wind, or it could be shock ionized gas (Binette \& Krongold 2008b), as a result of the fast wind striking the surrounding material.

Ton 34 provides evidence for a quasar where an accretion disk wind may be producing most of the emission and absorption features present in the UV-Optical spectra. Further observations of this source, and the associated evolution of the absorbing properties of its wind, hold strong diagnostic power to better understand the structure and geometry of quasar outflows.

\acknowledgments

We thank the anonymous referee for his/her constructive suggestions. Y.K. acknowledges support from the grant PAIIPIT IN104215 and CONACYT grant168519. L.Bianchi acknowledges support for Program number HST-GO-12510, provided by NASA through a grant from the Space Telescope Science Institute, which is operated by the
Association of Universities for Research in Astronomy, Incorporated, under NASA contract NAS5-26555.

\clearpage

\clearpage

\clearpage

\clearpage
\begin{deluxetable}{cccc}
\tablecolumns{4}
\tablewidth{0pc}
\tablecaption{Emission Lines in the HST-STIS Data}
\tablehead{
\colhead{Species}    & \colhead{$\lambda$(\AA)} &     \colhead{FWHM} &   \colhead{Flux} \\
\colhead{} & \colhead{}   & \colhead{(km s$^{-1}$)}   & \colhead{($10^{-14}$ erg s$^{-1}$ cm$^{-2}$)}}
\startdata
\cutinhead{Broad Components}
C IV  &  1549    &   5230 &  4.9  \\
Si IV  &  1401   &   --       &  2.9  \\
C II   &   1336   &   --       &  0.3  \\
Si II  &  1306   &   --       &  1.7  \\
Si II  &  1263   &   --       &  4.5  \\
N V  &  1240    &   --        &  10.9 \\
Ly$\alpha$ &  1216   &   --        &  24.5 \\
Si II  &  1193   &   --       &  4.4  \\
C III*   &  1175   &   --       &  1.1  \\
O VI &  1035   &   --     &  5.0 \\
Ly$\beta$ &  1026   &   --        &  1.7 \\
\cutinhead{Narrow Components}
C IV &  1549   &   1342   &  0.13 \\
Si IV &  1400   &   --       &   0.13 \\
Ly$\alpha$ &  1216   &   --     &  1.8
\enddata
\end{deluxetable}


\begin{thebibliography}{}

\bibitem[Andrade-Vel{\'a}zquez et al.(2010)]{2010ApJ...711..888A} Andrade-Vel{\'a}zquez, M., Krongold, Y., Elvis, M., et al.\ 2010, \apj, 711, 888 
\bibitem[Arav et al.(2001)]{2001ApJ...561..118A} Arav, N., de Kool, M., Korista, K.~T., et al.\ 2001, \apj, 561, 118 
\bibitem[Arav et al.(2013)]{2013MNRAS.436.3286A} Arav, N., Borguet, B., Chamberlain, C., Edmonds, D., \& Danforth, C.\ 2013, \mnras, 436, 3286 
\bibitem[Baldwin et al.(1996)]{1996ApJ...461..664B} Baldwin, J.~A., Ferland, G.~J., Korista, K.~T., et al.\ 1996, \apj, 461, 664 
\bibitem[Baskin \& Laor(2012)]{2012MNRAS.426.1144B} Baskin, A., \& Laor, A.\ 2012, \mnras, 426, 1144 
\bibitem[Baskin et al.(2013)]{2013MNRAS.432.1525B} Baskin, A., Laor, A., \& Hamann, F.\ 2013, \mnras, 432, 1525 
\bibitem[Baskin et al.(2015)]{2015MNRAS.449.1593B} Baskin, A., Laor, A., \& Hamann, F.\ 2015, \mnras, 449, 1593 
\bibitem[Barlow(1994)]{1994PASP..106..548B} Barlow, T.~A.\ 1994, \pasp, 106, 548 
\bibitem[Becker et al.(2000)]{2000ApJ...538...72B} Becker, R.~H., White, R.~L., Gregg, M.~D., et al.\ 2000, \apj, 538, 72 
\bibitem[Binette et al.(2005)]{2005ApJ...631..661B} Binette, L., Magris C., G., Krongold, Y., et al.\ 2005, \apj, 631, 661 
\bibitem[Binette \& Krongold(2008)]{2008A&A...477..413B} Binette, L., \& Krongold, Y.\ 2008, \aap, 477, 413 
\bibitem[Binette \& Krongold(2008)]{2008A&A...478..739B} Binette, L., \& Krongold, Y.\ 2008, \aap, 478, 739 
\bibitem[Bohlin \& Hartig(1998)]{1998stis.rept...20B} Bohlin, R., \& Hartig, G.\ 1998, Space Telescope STIS Instrument Science Report  
\bibitem[Borguet et al.(2013)]{2013ApJ...762...49B} Borguet, B.~C.~J., Arav, N., Edmonds, D., Chamberlain, C., \& Benn, C.\ 2013, \apj, 762, 49 
\bibitem[Capellupo et al.(2009)]{2009AAS...21360803C} Capellupo, D.~M., Hamann, F., Shields, J.~C., Barlow, T.~A., \& Rodriguez Hidalgo, P.\ 2009, American Astronomical Society Meeting Abstracts \#213, 213, 608.03 
\bibitem[Capellupo et al.(2011)]{2011MNRAS.413..908C} Capellupo, D.~M., Hamann, F., Shields, J.~C., Rodr{\'{\i}}guez Hidalgo, P., \& Barlow, T.~A.\ 2011, \mnras, 413, 908 
\bibitem[Capellupo et al.(2012)]{2012MNRAS.422.3249C} Capellupo, D.~M., Hamann, F., Shields, J.~C., Rodr{\'{\i}}guez Hidalgo, P., \& Barlow, T.~A.\ 2012, \mnras, 422, 3249 
\bibitem[Capellupo et al.(2014)]{2014MNRAS.444.1893C} Capellupo, D.~M., Hamann, F., \& Barlow, T.~A.\ 2014, \mnras, 444, 1893 
\bibitem[Chen \& Qin(2015)]{2015ApJ...799...63C} Chen, Z.-F., \& Qin, Y.-P.\ 2015, \apj, 799, 63 
\bibitem[Chen et al.(2013)]{2013MNRAS.434.3275C} Chen, Z.-F., Li, M.-S., Huang, W.-R., Pan, C.-J., \& Li, Y.-B.\ 2013, \mnras, 434, 3275 
\bibitem[D'Elia et al.(2009)]{2009A&A...503..437D} D'Elia, V., Fiore, F., Perna, R., et al.\ 2009, \aap, 503, 437 
\bibitem[D'Elia et al.(2009)]{2009AIPC.1111..495D} D'Elia, V., Fiore, F., Nicastro, F., Perna, R., \& Krongold, Y.\ 2009, American Institute of Physics Conference Series, 1111, 495 
\bibitem[D'Elia et al.(2009)]{2009ApJ...694..332D} D'Elia, V., Fiore, F., Perna, R., et al.\ 2009, \apj, 694, 332 
\bibitem[Di Matteo et al.(2005)]{2005Natur.433..604D} Di Matteo, T., Springel, V., \& Hernquist, L.\ 2005, \nat, 433, 604 
\bibitem[Elvis(2000)]{2000ApJ...545...63E} Elvis, M.\ 2000, \apj, 545, 63 
\bibitem[Dunn et al.(2010)]{2010ApJ...709..611D} Dunn, J.~P., Bautista, M., Arav, N., et al.\ 2010, \apj, 709, 611 
\bibitem[Fields et al.(2005)]{2005ApJ...634..928F} Fields, D.~L., Mathur, S., Pogge, R.~W., et al.\ 2005, \apj, 634, 928 
\bibitem[Filiz Ak et al.(2012)]{2012ApJ...757..114F} Filiz Ak, N., Brandt, W.~N., Hall, P.~B., et al.\ 2012, \apj, 757, 114 
\bibitem[Filiz Ak et al.(2013)]{2013ApJ...777..168F} Filiz Ak, N., Brandt, W.~N., Hall, P.~B., et al.\ 2013, \apj, 777, 168 
\bibitem[Gabel et al.(2005)]{2005ApJ...623...85G} Gabel, J.~R., Arav, N., Kaastra, J.~S., et al.\ 2005, \apj, 623, 85 
\bibitem[Gabel et al.(2005)]{2005ApJ...631..741G} Gabel, J.~R., Kraemer, S.~B., Crenshaw, D.~M., et al.\ 2005, \apj, 631, 741 
\bibitem[Gabel et al.(2005)]{2005AAS...207.1809G} Gabel, J.~R., Arav, N., \& Kim, T.-S.\ 2005, Bulletin of the American Astronomical Society, 37, 18.09 
\bibitem[Gipson et al.(2008)]{2008mefu.conf..157G} Gipson, J., MacMillan, D., \& Petrov, L.\ 2008, Measuring the Future, Proceedings of the Fifth IVS, 157 
\bibitem[Gipson \& Ray(2009)]{2009EGUGA..1113096G} Gipson, J.~M., \& Ray, R.~D.\ 2009, EGU General Assembly Conference Abstracts, 11, 13096 
\bibitem[Gipson et al.(2010)]{2010ivs..conf...90G} Gipson, J., Behrend, D., Gordon, D., et al.\ 2010, Sixth International VLBI Service for Geodesy and Astronomy.~ Proceedings from the 2010 General Meeting, ''VLBI2010: From Vision to Reality''.~Held 7-13 February, 2010 in Hobart, Tasmania, Australia.~Edited by D.~Behrend and K.D.~Baver.~ NASA/CP 2010-215864., p.90-94, 90 
\bibitem[Giustini(2015)]{2015ebha.confE..32G} Giustini, M.\ 2015, The Extremes of Black Hole Accretion, 32 
\bibitem[Goodrich \& Miller(1995)]{1995ApJ...448L..73G} Goodrich, R.~W., \& Miller, J.~S.\ 1995, \apjl, 448, L73 
\bibitem[Grevesse \& Sauval(1998)]{1998SSRv...85..161G} Grevesse, N., \& Sauval, A.~J.\ 1998, \ssr, 85, 161 
\bibitem[Gupta et al.(2013)]{2013ApJ...768..141G} Gupta, A., Mathur, S., Krongold, Y., \& Nicastro, F.\ 2013, \apj, 768, 141 
\bibitem[Gupta et al.(2013)]{2013ApJ...772...66G} Gupta, A., Mathur, S., Krongold, Y., \& Nicastro, F.\ 2013, \apj, 772, 66 
\bibitem[Gupta et al.(2015)]{2015ApJ...798....4G} Gupta, A., Mathur, S., \& Krongold, Y.\ 2015, \apj, 798, 4 
\bibitem[Hall et al.(2011)]{2011MNRAS.411.2653H} Hall, P.~B., Anosov, K., White, R.~L., et al.\ 2011, \mnras, 411, 2653 
\bibitem[Hamann \& Sabra(2004)]{2004ASPC..311..203H} Hamann, F., \& Sabra, B.\ 2004, AGN Physics with the Sloan Digital Sky Survey, 311, 203 
\bibitem[Hamann et al.(2008)]{2008MNRAS.391L..39H} Hamann, F., Kaplan, K.~F., Rodr{\'{\i}}guez Hidalgo, P., Prochaska, J.~X., \& Herbert-Fort, S.\ 2008, \mnras, 391, L39 
\bibitem[Hamann et al.(2011)]{2011MNRAS.410.1957H} Hamann, F., Kanekar, N., Prochaska, J.~X., et al.\ 2011, \mnras, 410, 1957 
\bibitem[Hamann et al.(2012)]{2012ASPC..460...47H} Hamann, F., Simon, L., Rodriguez Hidalgo, P., \& Capellupo, D.\ 2012, AGN Winds in Charleston, 460, 47 
\bibitem[Hamann et al.(2013)]{2013MNRAS.435..133H} Hamann, F., Chartas, G., McGraw, S., et al.\ 2013, \mnras, 435, 133 
\bibitem[Haro-Corzo et al.(2007)]{2007ApJ...662..145H} Haro-Corzo, S.~A.~R., Binette, L., Krongold, Y., et al.\ 2007, \apj, 662, 145 
\bibitem[Hazard et al.(1984)]{1984ApJ...282...33H} Hazard, C., Morton, D.~C., Terlevich, R., \& McMahon, R.\ 1984, \apj, 282, 33 
\bibitem[Hines \& Wills(1995)]{1995ApJ...448L..69H} Hines, D.~C., \& Wills, B.~J.\ 1995, \apjl, 448, L69 
\bibitem[Hopkins et al.(2005)]{2005ApJ...630..705H} Hopkins, P.~F., Hernquist, L., Cox, T.~J., et al.\ 2005, \apj, 630, 705 
\bibitem[Junkkarinen et al.(2001)]{2001AAS...198.7401J} Junkkarinen, V.~T., Cohen, R.~D., Hamann, F., \& Shields, G.~A.\ 2001, Bulletin of the American Astronomical Society, 33, 74.01 
\bibitem[Kaastra et al.(2014)]{2014Sci...345...64K} Kaastra, J.~S., Kriss, G.~A., Cappi, M., et al.\ 2014, Science, 345, 64 
\bibitem[Kaspi et al.(2007)]{2007ApJ...659..997K} Kaspi, S., Brandt, W.~N., Maoz, D., et al.\ 2007, \apj, 659, 997 
\bibitem[King(2010)]{2010ASPC..427..315K} King, A.~R.\ 2010, Accretion and Ejection in AGN: a Global View, 427, 315 
\bibitem[Kormendy(2013)]{2013IAUS..295..241K} Kormendy, J.\ 2013, The Intriguing Life of Massive Galaxies, 295, 241 
\bibitem[Kraemer et al.(2003)]{2003ApJ...582..125K} Kraemer, S.~B., Crenshaw, D.~M., Yaqoob, T., et al.\ 2003, \apj, 582, 125 
\bibitem[Krongold et al.(2003)]{2003ApJ...597..832K} Krongold, Y., Nicastro, F., Brickhouse, N.~S., et al.\ 2003, \apj, 597, 832 
\bibitem[Krongold et al.(2005)]{2005ApJ...620..165K} Krongold, Y., Nicastro, F., Elvis, M., et al.\ 2005, \apj, 620, 165 
\bibitem[Krongold et al.(2005)]{2005ApJ...622..842K} Krongold, Y., Nicastro, F., Brickhouse, N.~S., Elvis, M., \& Mathur, S.\ 2005, \apj, 622, 842 
\bibitem[Krongold et al.(2007)]{2007ApJ...659.1022K} Krongold, Y., Nicastro, F., Elvis, M., et al.\ 2007, \apj, 659, 1022 
\bibitem[Krongold et al.(2009)]{2009ApJ...690..773K} Krongold, Y., Jim{\'e}nez-Bail{\'o}n, E., Santos-Lleo, M., et al.\ 2009, \apj, 690, 773 
\bibitem[Krongold et al.(2010)]{2010ApJ...724L.203K} Krongold, Y., Binette, L., \& Hern{\'a}ndez-Ibarra, F.\ 2010, \apjl, 724, L203 
\bibitem[Krongold \& Prochaska(2013)]{2013ApJ...774..115K} Krongold, Y., \& Prochaska, J.~X.\ 2013, \apj, 774, 115 
\bibitem[Laor et al.(1997)]{1997ApJ...489..656L} Laor, A., Jannuzi, B.~T., Green, R.~F., \& Boroson, T.~A.\ 1997, \apj, 489, 656 
\bibitem[Leighly et al.(2009)]{2009ApJ...701..176L} Leighly, K.~M., Hamann, F., Casebeer, D.~A., \& Grupe, D.\ 2009, \apj, 701, 176 
\bibitem[Leighly et al.(2015)]{2015ApJ...809L..13L} Leighly, K.~M., Cooper, E., Grupe, D., Terndrup, D.~M., \& Komossa, S.\ 2015, \apjl, 809, L13 
\bibitem[Leighly et al.(2007)]{2007ApJ...663..103L} Leighly, K.~M., Halpern, J.~P., Jenkins, E.~B., et al.\ 2007, \apj, 663, 103 
\bibitem[Longinotti et al.(2013)]{2013ApJ...766..104L} Longinotti, A.~L., Krongold, Y., Kriss, G.~A., et al.\ 2013, \apj, 766, 104 
\bibitem[Longinotti et al.(2015)]{2015ApJ...813L..39L} Longinotti, A.~L., Krongold, Y., Guainazzi, M., et al.\ 2015, \apjl, 813, L39 
\bibitem[Lundgren et al.(2007)]{2007ApJ...656...73L} Lundgren, B.~F., Wilhite, B.~C., Brunner, R.~J., et al.\ 2007, \apj, 656, 73 
\bibitem[Ma(2002)]{2002MNRAS.335L..99M} Ma, F.\ 2002, \mnras, 335, L99 
\bibitem[McGraw et al.(2015)]{2015MNRAS.453.1379M} McGraw, S.~M., Shields, J.~C., Hamann, F.~W., et al.\ 2015, \mnras, 453, 1379 
\bibitem[Maiolino et al.(2010)]{2010Msngr.142...36M} Maiolino, R., Mannucci, F., Cresci, G., et al.\ 2010, The Messenger, 142, 36 
\bibitem[Mathur et al.(1995)]{1995ApJ...455L...9M} Mathur, S., Elvis, M., \& Singh, K.~P.\ 1995, \apjl, 455, L9 
\bibitem[Matthews et al.(2016)]{2016MNRAS.458..293M} Matthews, J.~H., Knigge, C., Long, K.~S., et al.\ 2016, \mnras, 458, 293 
\bibitem[Miniutti et al.(2014)]{2014MNRAS.437.1776M} Miniutti, G., Sanfrutos, M., Beuchert, T., et al.\ 2014, \mnras, 437, 1776 
\bibitem[Moe et al.(2009)]{2009ApJ...706..525M} Moe, M., Arav, N., Bautista, M.~A., \& Korista, K.~T.\ 2009, \apj, 706, 525 
\bibitem[Murray et al.(1995)]{1995ApJ...451..498M} Murray, N., Chiang, J., Grossman, S.~A., \& Voit, G.~M.\ 1995, \apj, 451, 498 
\bibitem[Misawa et al.(2014)]{2014ApJ...792...77M} Misawa, T., Charlton, J.~C., \& Eracleous, M.\ 2014, \apj, 792, 77 
\bibitem[Misawa et al.(2014)]{2014ApJ...794L..20M} Misawa, T., Inada, N., Oguri, M., et al.\ 2014, \apjl, 794, L20 
\bibitem[Nardini et al.(2015)]{2015Sci...347..860N} Nardini, E., Reeves, J.~N., Gofford, J., et al.\ 2015, Science, 347, 860 
\bibitem[Nardini et al.(2015)]{2015MNRAS.453.2558N} Nardini, E., Gofford, J., Reeves, J.~N., et al.\ 2015, \mnras, 453, 2558 
\bibitem[Nicastro et al.(1999)]{1999ApJ...512..184N} Nicastro, F., Fiore, F., Perola, G.~C., \& Elvis, M.\ 1999, \apj, 512, 184 
\bibitem[Ostriker et al.(2010)]{2010ApJ...722..642O} Ostriker, J.~P., Choi, E., Ciotti, L., Novak, G.~S., \& Proga, D.\ 2010, \apj, 722, 642 
\bibitem[Proga et al.(2012)]{2012ASPC..460..171P} Proga, D., Rodriguez-Hidalgo, P., \& Hamann, F.\ 2012, AGN Winds in Charleston, 460, 171 
\bibitem[Rodr{\'{\i}}guez Hidalgo et al.(2011)]{2011MNRAS.411..247R} Rodr{\'{\i}}guez Hidalgo, P., Hamann, F., \& Hall, P.\ 2011, \mnras, 411, 247 
\bibitem[Rodr{\'{\i}}guez Hidalgo et al.(2013)]{2013ApJ...775...14R} Rodr{\'{\i}}guez Hidalgo, P., Eracleous, M., Charlton, J., et al.\ 2013, \apj, 775, 14 
\bibitem[Risaliti et al.(2011)]{2011MNRAS.410.1027R} Risaliti, G., Nardini, E., Salvati, M., et al.\ 2011, \mnras, 410, 1027 
\bibitem[Sanfrutos et al.(2016)]{2016MNRAS.457..510S} Sanfrutos, M., Miniutti, G., Krongold, Y., Ag{\'{\i}}s-Gonz{\'a}lez, B., \& Longinotti, A.~L.\ 2016, \mnras, 457, 510 
\bibitem[Sargent et al.(1988)]{1988ApJS...68..539S} Sargent, W.~L.~W., Boksenberg, A., \& Steidel, C.~C.\ 1988, \apjs, 68, 539 
\bibitem[Sargent et al.(1988)]{1988ApJ...334...22S} Sargent, W.~L.~W., Steidel, C.~C., \& Boksenberg, A.\ 1988, \apj, 334, 22 
\bibitem[Suzuki(2006)]{2006ApJS..163..110S} Suzuki, N.\ 2006, \apjs, 163, 110 
\bibitem[Scannapieco \& Oh(2004)]{2004ApJ...608...62S} Scannapieco, E., \& Oh, S.~P.\ 2004, \apj, 608, 62 
\bibitem[Silk(2011)]{2011IAUS..277..273S} Silk, J.\ 2011, Tracing the Ancestry of Galaxies, 277, 273 
\bibitem[Tombesi et al.(2015)]{2015Natur.519..436T} Tombesi, F., Mel{\'e}ndez, M., Veilleux, S., et al.\ 2015, \nat, 519, 436 
\bibitem[Vivek et al.(2012)]{2012MNRAS.421L.107V} Vivek, M., Srianand, R., Mahabal, A., \& Kuriakose, V.~C.\ 2012, \mnras, 421, L107 
\bibitem[Vivek et al.(2012)]{2012MNRAS.423.2879V} Vivek, M., Srianand, R., Petitjean, P., et al.\ 2012, \mnras, 423, 2879 
\bibitem[Vivek et al.(2014)]{2014MNRAS.440..799V} Vivek, M., Srianand, R., Petitjean, P., et al.\ 2014, \mnras, 440, 799 
\bibitem[Vivek et al.(2015)]{2015yCat..74400799V} Vivek, M., Srianand, R., Petitjean, P., et al.\ 2015, VizieR Online Data Catalog, 744,  
\bibitem[Vivek et al.(2016)]{2016MNRAS.455..136V} Vivek, M., Srianand, R., \& Gupta, N.\ 2016, \mnras, 455, 136 
\bibitem[Wang et al.(2015)]{2015ApJ...814..150W} Wang, T., Yang, C., Wang, H., \& Ferland, G.\ 2015, \apj, 814, 150 
\bibitem[Welling et al.(2014)]{2014MNRAS.440.2474W} Welling, C.~A., Miller, B.~P., Brandt, W.~N., Capellupo, D.~M., \& Gibson, R.~R.\ 2014, \mnras, 440, 2474 
\bibitem[Weymann et al.(1991)]{1991ApJ...373...23W} Weymann, R.~J., Morris, S.~L., Foltz, C.~B., \& Hewett, P.~C.\ 1991, \apj, 373, 23 

\end{thebibliography}
\end{document}